# The dynamics of planetary nebulae in the Galaxy
## Evidence for a third integral.

S. Durand[1,2], H. Dejonghe[2], and A. Acker[1]

[1] URA 1280, Equipe évolution galactique, Observatoire de Strasbourg, 11 rue de l'Université, F–67000 Strasbourg, France
[2] Universiteit Gent, Sterrenkundig Observatorium, Krijgslaan 281, B–9000 Gent, Belgium

**Abstract.** We present a dynamical analysis of 673 galactic Planetary Nebulae, using a two-integral axisymmetric model with a Kuzmin-Kutuzov Stäckel potential. The method fits the kinematics to the projected moments of a distribution function, by means of Quadratic Programming. The 2.2 $\mu$m COBE brightness map has been used after correction for the interstellar extinction as a projected star counts map in the modeling, because it constitutes a galactic distribution view of evolved red populations which are considered to be the progenitors of PNe.
The model we have obtained provides a 2-integral distribution function for the COBE 2.2 $\mu$m map, and thus *a fortiori* a deprojection of it, which allows moreover the identification of all the major Galactic components. We derive the density laws for them. The projected velocity dispersions are not well fitted though, especially in the disk, which points at the likely presence of a third integral. If this result can be confirmed by additional data, this would mean that for the first time the presence and importance of a third integral on a global scale is demonstrated.

## 1. Introduction

The dynamical modeling of various stellar populations in our Galaxy is an important tool in the study of its formation and evolution. Any such sample of course must consist of selected candidates, which often restrict our attention to luminous stars such as K- and M- Giants (Terndrup 1993), variable stars such as Mira's (Whitelock, 1993), post-AGB stars such as OH/IR stars (Sevenster, Dejonghe & Habing 1995) or Planetary Nebulae, hereafter PNe (Aller, 1993). None of these tracer objects necessarily belong to the main-stream pack, which complicates the interpretation of the data.
Dynamical studies of the Galaxy as a whole need reasonably homogeneous samples over large volumes. At present, only the OH/IR stars (te Lintel Hekkert, 1990) and the PNe come into consideration for this. The OH/IR stars, which are strong infrared emittors, oxygen-rich post-AGB stars in the final state of the red giant phase are especially suited, since the available samples are products of blind surveys or result from well-defined selection procedures on otherwise blind surveys (notably IRAS), as demonstrated by Pottasch et al. (1988) and Ratag et al. (1990). As to the PNe, they are transient objects of considerable astrophysical interest. Their existence is based upon the synchronous evolution of their 2 components : a hot central star surrounded by a circumstellar nebula in expansion. PNe are thought to descend from low- and intermediate-mass stars ($M_i < 8M_\odot$) and probably cover a large range of ages. They represent a very short and very late phase in stellar evolution, located beyond the AGB stage. In their death agony they probably end up as white dwarfs. PNe can be tracked down relatively easily throughout the entire galaxy thanks to their rich emission line spectra that make them suitable as test particles for dynamical studies.

## 2. The database

Up to now, 1143 objects are classified as "true or probable" PNe in the *Strasbourg-ESO Catalogue of Galactic PNe* (Acker et al., 1992). Most of them seem to belong to a thick disk population, while, in accordance to the bulge selection criteria of the Catalogue ($|l|$ and $|b| < 10°$, radio fluxes $F(6cm) < 100\ mJy$ and angular diameters $< 20\ arcsec$), about 300 are probably bulge objects.
We will use only the 673 PNe for which radial velocities are available. The database used in our modeling comprises two distinct data sets :

*2.1. Galactic positions and radial velocities of 673 PNe*

The radial velocities stem from miscellaneous spectroscopic surveys. Most are compiled originally in Schneider & Terzian (1983) and in the Strasbourg-ESO Catalogue (Acker et al., 1992). Some others, about one hundred still non-published data with a precision of about 50 km/s, come from the bulge survey of low spectral resolution performed by A. Acker and B. Stenholm (Acker et al., 1991). Moreover, about 60 radial velocities with a precision of nearly 1 km/s and originated from high resolution observations performed by A. Acker and A. Zijlstra were added to our sample. These last radial velocities constituted the 'reference set' for the calibration of the low- and medium-resolution ones (Durand et al., 1995).

*Send offprint requests to*: S. Durand

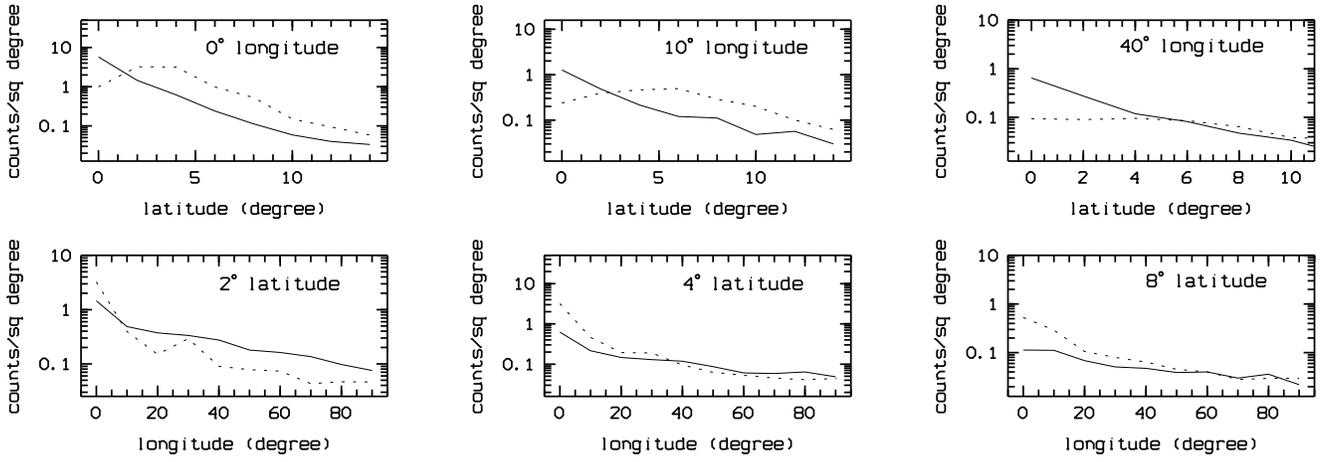

**Fig. 2.** A comparison of PNe (dotted curves) with the extinction corrected intensities provided by COBE at 2.2 μm (solid curves), the latter scaled so as to produce roughly the same numbers as the other two. Note the effects of the (optical) extinction in the PNe sample at small latitude, and the positive detection bias for Bulge PNe.

About 300 of these 673 PNe are thought to belong to the galactic bulge.

Since the Strasbourg-ESO catalog is to a large extent a compilation of all known or conjectured PNe in the literature, its selection biases are rather poorly understood. Moreover, the catalog draws mostly from optical work, and is thus severely incomplete in the plane. Therefore, it is not an easy-to-use instrument for producing star counts, which are affected by selection biases to zeroth-order and extinction effects. For example, we can easily identify a selection effect due to the special attention which the bulge attracts, causing a disproportional detection efficiency of PNe in that region (see also discussion of figure 2). We will thus use the catalog only to obtain the kinematics of the PNe, i.e. projected mean streaming and projected velocity dispersion.

## 2.2. COBE intensity map of the full sky at 2.2 μm

At 2.2 μm, emissions of late-K and M giants dominate the spectrum (Arendt et al., 1994). Our galaxy is furthermore relatively transparent to (cold) interstellar material in that K band, and thermal emission from dust is still negligible. Hence, the 2.2 μm COBE picture is a rather reliable map of the projected spatial distribution of galactic AGB stars. Moreover, there is at the present day a general "*consensus that AGB stars are immediate progenitors of PNe*" (Habing et al., 1993): red giants and PNe have evolutionary links through the OH/IR phase, although the precise nature of all the transition objects is not yet really known (Habing, 1990). So it is not unreasonable at first sight to use the 2.2 μm COBE map as a substitute for the spatial distributions of galactic AGB and post-AGB populations. We will now show that corroborating evidence for the validity of this procedure can be found in the comparison of the data sets.

Clearly the raw COBE map must be corrected for interstellar extinction. Following Arendt et al. (1994), we used the relations between near IR colours (caused by extinction), as determined by Rieke & Lebofsky (1985). We first constructed a colour excess map $E(J - K)$ by the combination of the two first near IR brightness COBE maps, and by choosing a unique intrinsic colour $(J - K)$ for the Galaxy, thus neglecting the possible presence of a variation in the intrinsic colour between the inner and the outer parts of the Galaxy. We have then built an optical depth map of the galaxy at 2.2 μm by associating the reddening law. Finally we transformed the 2.2 $\mu$m brightness COBE map into a counts density map by a simple scaling, so as to reproduce roughly the same numbers as the stars sample. Figure 2 compares the PNe counts and the scaled 2.2 $\mu$m COBE map. The first row of plots clearly shows the effect of the optical extinction, which affects the PNe counts. At larger latitudes however, the curves are very similar. At the present day it's hardly possible to correct for the extinction in these stars counts. The second row reveals that, even in the presence of extinction at $b = 2°$, the slopes remain very similar for $|l| > 20°$. On the other hand, our PNe counts appear effectively overestimated in the bulge, and that can be well understood since we have added to the original PNe catalog nearly a hundred of new bulge measurements.

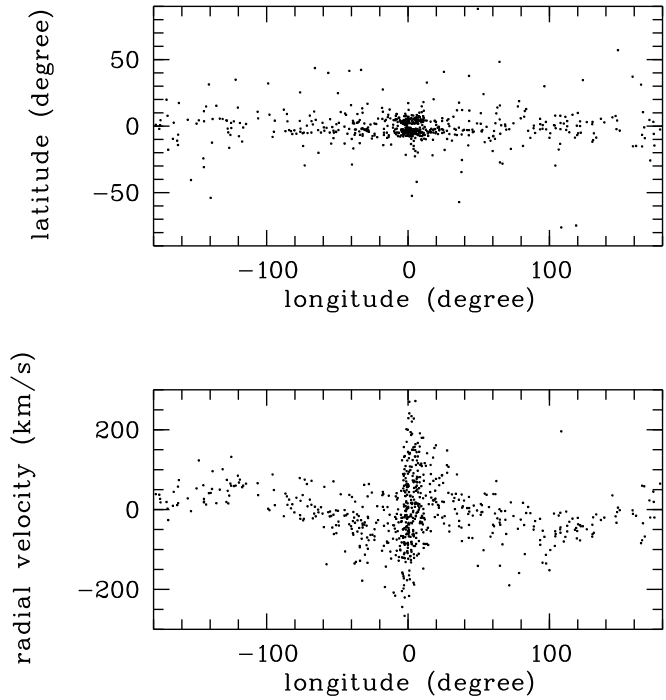

**Fig. 1.** The galactic distribution (up) and the radial velocities versus galactic longitude (down) of our 673 PNe sample.

From this we see that it is certainly reasonable to use the COBE map as a substitute for the PNe counts, in order to overcome the different biases in the PNe catalogs. This can be asserted, not only for general reasons of astrophysical nature, but also on the basis of the comparison of both data sets, there where it makes sense to compare them.

We have restricted our study to the galactic region $2 < |b| < 8$ for the following reasons:

- $|b| < 2$ is likely to be a very complicated region where optical depths reach their highest values, in so far as the extinction corrected COBE map cannot reasonably represent the true galactic K emission.
- PNe data at $|b| > 8$ are unfortunately not numerous enough to allow a correct estimation of the different moments of the velocity distribution.

Moreover, the contribution of the zodiacal light is minimal in the K band as evidenced in the composite 1.25, 2.2 and 3.5 $\mu$m COBE images (Hauser 1993), so we have neglected it as is common practice in near-infrared studies of the Galactic emissivity distribution (see also Kent et al., 1991). In our case in particular, this approximation is not critical, since the effect reaches levels that are below the accuracy we can reasonably obtain for the 0-th moment of the distribution function.

## 3. The modeling

### 3.1. The framework

PNe are highly evolved objects supposed to represent a dynamicaly relaxed population. Therefore, they are in a sufficiently steady state to legitimate the application of an equilibrium model. One of the first choices to be made is the modeling geometry: we chose here the axisymmetric approximation. This could appear at first sight as rather outdated, because of the mounting evidence that the Galaxy is a complex triaxial system (e.g. Spergel D.N., 1992). However, the triaxiality in PNe kinematics is not obvious, if not absent, so it seems legitimate to us (by Occam's razor) to firstly exploit the axisymmetric geometry.

Our basic aim is the construction of the *distribution function* of the PNe population, which is simply the probability density of the sample in the 6 dimensional phase space $(x, v)$. It will be later of great help to appreciate the different orbital structures present in the PNe population (section 4).

The potential we use in this work is a *Kuzmin-Kutuzov* Stäckel potential with a halo-disk structure (Batsleer & Dejonghe 1994). Its general expression in spheroidal coordinates can be written as

$$V(\lambda_d, \nu_d, q) = V(\lambda_d, \nu_d) + V(\lambda_h, \nu_h)$$
$$= -GM\left(\frac{K}{\sqrt{\lambda_d} + \sqrt{\nu_d}} + \frac{1-K}{\sqrt{\lambda_d - q} + \sqrt{\nu_d - q}}\right)$$

with :

- $d$ and $h$ standing for 'disk' and 'halo'
- $\lambda, \nu$ spheroidal coordinates
- $G$ the constant of gravity
- $M$ the total mass of the galaxy
- $K = M_d/M$
- $q$ a parameter defined by $q = \lambda_d - \lambda_h = \nu_d - \nu_h \geq 0$.

Let's recall briefly that spheroidal coordinates $(\lambda, \nu)$ are defined as the roots for $\tau$ of the equation :

$$\frac{r^2}{\tau - \alpha} + \frac{z^2}{\tau - \gamma} = 1$$

where $(r, \phi, z)$ are the cylindrical coordinates and $\alpha, \gamma$ negative constants. The axis ratios $\frac{a}{c}$ of each of the two coordinate surfaces (with $a = -\sqrt{\alpha}$ and $c = -\sqrt{\gamma}$) and the $K$ parameter are selected in order to reproduce the shape of the galactic rotation curve with its characteristic flat extent. For this, we use for our mass model:

$$\frac{a_d}{c_d} = 75, \quad \frac{a_h}{c_h} = 1.01, \quad K = 0.07, \quad M = 4.2 \times 10^{11} M_\odot$$

The selection of the parameters is based upon the requirements that the rotation curve be flat, that the radial and thickness scalelengths of the disk are close to the commonly accepted values, and that the surface and spatial densities of the disk in the solar neighbourhood is reproduced (for more details about the selection of the parameters, see Batsleer & Dejonghe, 1994). Note that the shape of the halo is almost spherical and that the adopted $K$ parameter implies a disk mass of $2.94 \times 10^{10} M_\odot$.

The (stationary) distribution function is usually described via Jeans' Theorem as a function of *integrals of motion* which are, in the present case, the orbital binding energy $E$, the $z$-component of the angular momentum $L_z$, and a third integral which is simple and analytic in the Stäckel case. Since information about vertical motions is still practically non-existent, the explicit presence of a third integral in the distribution function will not directly reveal itself in the kinematics. Therefore, only the failure of a complete dynamical 2-integral model to fit the data may point to the presence of the third integral. The major approximation made with the utilization of a *two-integral* model is that radial and vertical dispersions are supposed to be equal; this can be easily checked just by looking at the expressions, in cylindrical coordinates, of the two first integrals of motion $E = \psi(r,z) - \frac{1}{2}(v_r^2 + v_\theta^2 + v_z^2)$ and $L_z = r\, v_\theta$. The contributions of the $v_r$ and $v_z$ components are completely identical contrary to the $v_\theta$ component which appears in an independent way in $L_z$. That 2-integral approximation should be rather good in the thin disk where dispersions are thought to be lower than anywhere else since orbits there are nearly circular, but problems should arise elsewhere, and more particularly in the thick disk.

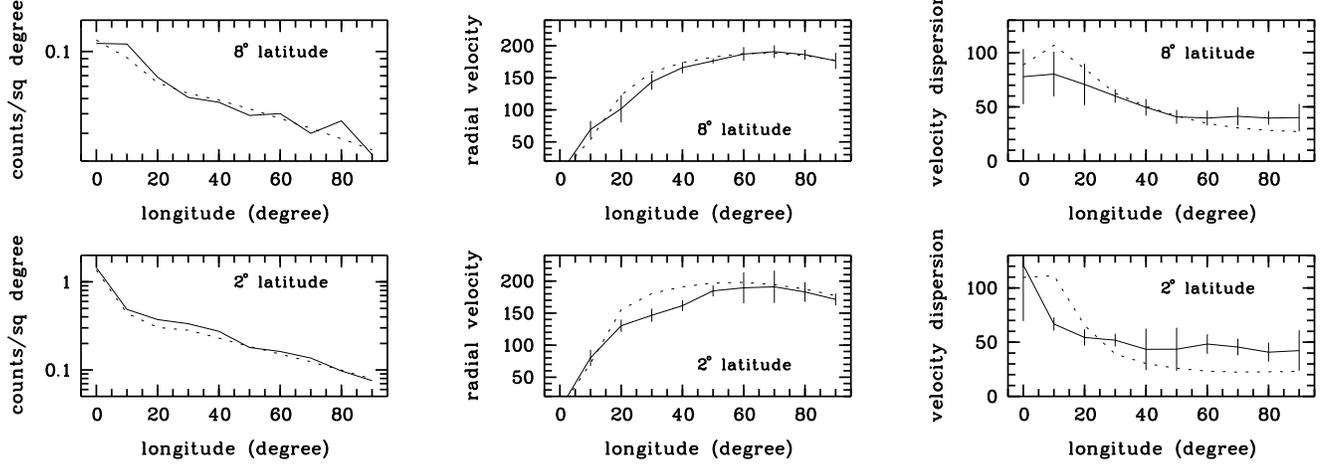

**Fig. 3.** A comparison of the data (solid curves) and of our best 2 integral model (dashed curves) at different galactic positions. From left to right : the projected counts per square degree, the projected mean velocities (in km/s) and the projected velocity dispersions (in km/s). The error bars on the PNe kinematics have been obtained by varying the sampling strategy.

### 3.2. The QP method

In this method, the distribution function is written as a sum of elementary distribution functions,

$$F(E, L_z) = \sum_i c_i F_i(E, L_z) \quad \text{with } i = i(\alpha, \beta, \gamma), \quad (1)$$

which are of 2 kinds (families) that we call 'building blocks':

1. $F_{\alpha,\beta}(E, L) = E^\alpha (E \frac{L^2}{2})^\beta$,
   which are essentially Fricke components associated with fairly slowly rotating models with high velocity dispersion (halo-type).
2. $F_{\alpha,\beta,\gamma}(E, L) = exp(-\frac{\alpha_1}{S}) S^{\alpha_2} (2SL^2)^\beta (E - S)^\gamma \quad if \ S < E$
   $F_{\alpha,\beta,\gamma}(E, L) = 0 \quad if \ S > E$
   with $S$ the minimum orbital energy of an orbit which can bring a star with an associated angular momentum $L_z$ at a given height $z_0$.

The second building block produces models with much smaller scaleheights than the first one, and with nearly circular orbits (disk-type). Note that components with $\beta > 0$ have a zero central density, and that the presence of the exponential factor is useful to control the radial behaviour of the mass distribution in the disk (Batsleer & Dejonghe, 1995).

In order to determine the coefficients $c_i$ of the distribution function (1), we consider a grid in galactic coordinates on which data points are placed: each of the grid points is associated with the $0^{th}$, $1^{st}$ and $2^{nd}$ moments of the projected velocity distribution, calculated by taking into account the 15 stars closest to the grid point (see figure 4 and table 1).

This yields for each point on the sky 3 observables $\mu_{obs}(x, v)$ that can be estimated by means of a function $\mu(x, v)$ :

$$\mu_{mod}(x_l, v_l) = \int \mu(x, v) F(I) \, dx \, dv$$
$$= \sum_i c_i \int_{\tau_l} \mu_{mod}(x, v) F_i(I) \, dx \, dv = \sum_i c_i \mu_i(x_l, v_l).$$

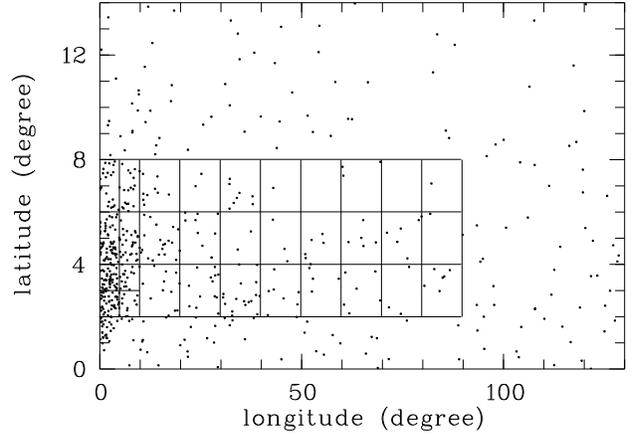

**Fig. 4.** The grid used for the evaluation of the different moments of the projected velocity distribution, superposed on nearly all our data points, with positions that were flipped into the first galactic quadrant.

In our case $\tau_l$ is the product of the full velocity space available at a spatial point and the line of sight. Then we can construct a $\chi^2$-type function :

$$\sum_l w_l [\mu_{obs}(x_l, v_l) - \sum_i c_i \mu_i(x_l, v_l)]^2,$$

with $w_l$ weights that can be freely chosen. This function is minimized during the calculations with additionnal constraints to force a positive distribution function in *all* phase space (Dejonghe, 1989). This is a *Quadratic Programming* problem. This method has the advantage to accomodate inhomogeneous data sets (photometry, counts, kinematics) and additionnaly imposed constraints, as long as these can be put in a form which is linear in the distribution function. Further details of the method as applied on similar problems can be found in Bat-sleer & Dejonghe (1995) and Sevenster, Dejonghe & Habing (1995).

Table 1. counts (in $10^{-2}$ counts/sq degree) providing by COBE, projected mean velocities (in km/s) and projected mean dispersions (in km/s) coming from PNe data, all given for each grid points (l,b).

| b | | l | 0 | 10 | 20 | 30 | 40 | 50 | 60 | 70 | 80 | 90 |
|---|---|---|---|---|---|---|---|---|---|---|---|---|
| | $\rho$ | | 145 | 49 | 37 | 33 | 27 | 18 | 16 | 14 | 10 | 8 |
| | $v$ | 2 | — | 80 | 130 | 147 | 162 | 185 | 190 | 191 | 183 | 172 |
| | $\sigma$ | | 121 | 67 | 54 | 52 | 43 | 43 | 48 | 45 | 41 | 42 |
| | $\rho$ | | 62 | 21 | 15 | 13 | 12 | 9 | 6 | 6 | 6 | 5 |
| | $v$ | 4 | — | 76 | 124 | 146 | 162 | 182 | 190 | 190 | 181 | 170 |
| | $\sigma$ | | 116 | 82 | 58 | 51 | 46 | 47 | 47 | 44 | 42 | 45 |
| | $\rho$ | | 24 | 12 | 9 | 8 | 8 | 6 | 7 | 5 | 4 | 3 |
| | $v$ | 6 | — | 59 | 102 | 152 | 162 | 179 | 189 | 187 | 182 | 171 |
| | $\sigma$ | | 73 | 82 | 67 | 51 | 44 | 46 | 48 | 44 | 39 | 44 |
| | $\rho$ | | 11 | 11 | 7 | 5 | 5 | 4 | 4 | 3 | 4 | 2 |
| | $v$ | 8 | — | 69 | 102 | 144 | 166 | 176 | 187 | 191 | 186 | 176 |
| | $\sigma$ | | 78 | 80 | 71 | 60 | 50 | 41 | 40 | 41 | 40 | 40 |

## 4. Results on PNe dynamics

### 4.1. The validity of the modeling

The quality of the fit can be checked directly by comparing the data and the model, as shown in figure 3 for different galactic regions. We see that the model fits perfectly the COBE density and the projected mean velocities. On the other hand, there is an obvious discrepancy in the fit of the projected dispersions, particularly in the disk where the errors are nearly of the same scale as the data : this is probably the signature of an explicit third integral dependence in the distribution function (see 3.1).

### 4.2. The orbital structures

#### 4.2.1. The distribution function in integral space

A schematic representation is shown in figure 5 and the PNe application in figure 8 (top right). Both figures are shown in the coordinate space $(L_z\sqrt{2E}, E)$, and not $(L_z, E)$ in order to keep integral space finite. For a given point $(r, z)$, the region of the allowed orbits is bounded by an ellipse which is tangential in two points at the curved outer boundaries which are the loci of the circular orbits (one point for prograde rotating populations, at $L_z > 0$, and another for retrograde rotating populations, at $L_z < 0$). We recall that the positive $L_z$ axis points towards the galactic north pole by convention, so our Galaxy is rotating retrograde. Each circular orbit curve (which occurs by definition at z=0) is shown as far as $r = 15$ kpc.

#### 4.2.2. The distribution function in turning point space

This representation has the advantage that it is more intuitive than the previous one. A schematic view is shown in figure 6 and the PNe application in figure 8 (top left). The coordinate system is $(r_p, r_a)$, being respectively the pericenter and the apocenter of the orbit. The loci of the circular orbits are defined by the lines $r_p = +r_a$ (prograde orbits) and $r_p = -r_a$ (retrograde orbits), and the allowed orbits are thereby placed

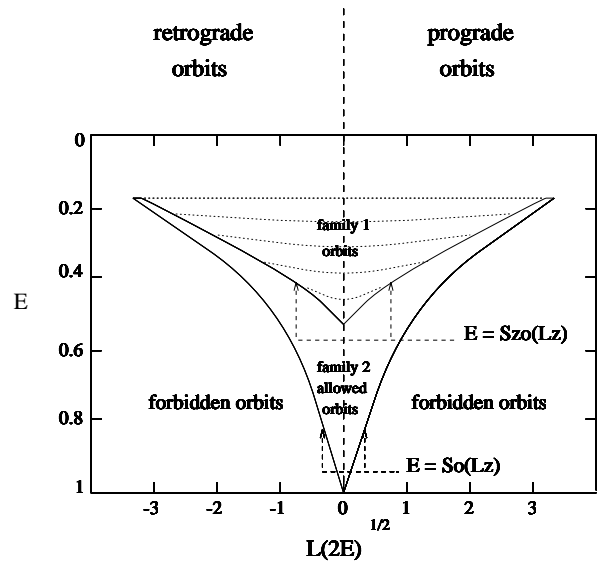

**Fig. 5.** A schematic representation of the distribution function in integral space $(L_z\sqrt{2E}, E)$ displayed up to a distance $r$ of 15 kpc from the galactic centre. The two curves $E = S_0(L_z)$ correspond to the circular orbits (occuring by definition at $z = 0$). The two curves $E = S_{z_0}(L_z)$ correspond to the minimum binding energy that can bring a star at most as high as $z_0$.

in the regions where $r_a \geq |r_p|$ ; by convention $r_p$ has the sign of $L_z$.

The 2 plates at the top of figure 8, presented in logarithmic density scale, show two very beautiful distribution functions which are quasi dominated by the second building-block family (disk-type). We note that most of the orbits are nearly circular in the disk, that the bulge is not excessively populated (the orbits are very dispersed), and that the phase space density of the old disk increases from $r = 5$ kpc to about $r = 10$ kpc, whereafter it decreases smoothly.

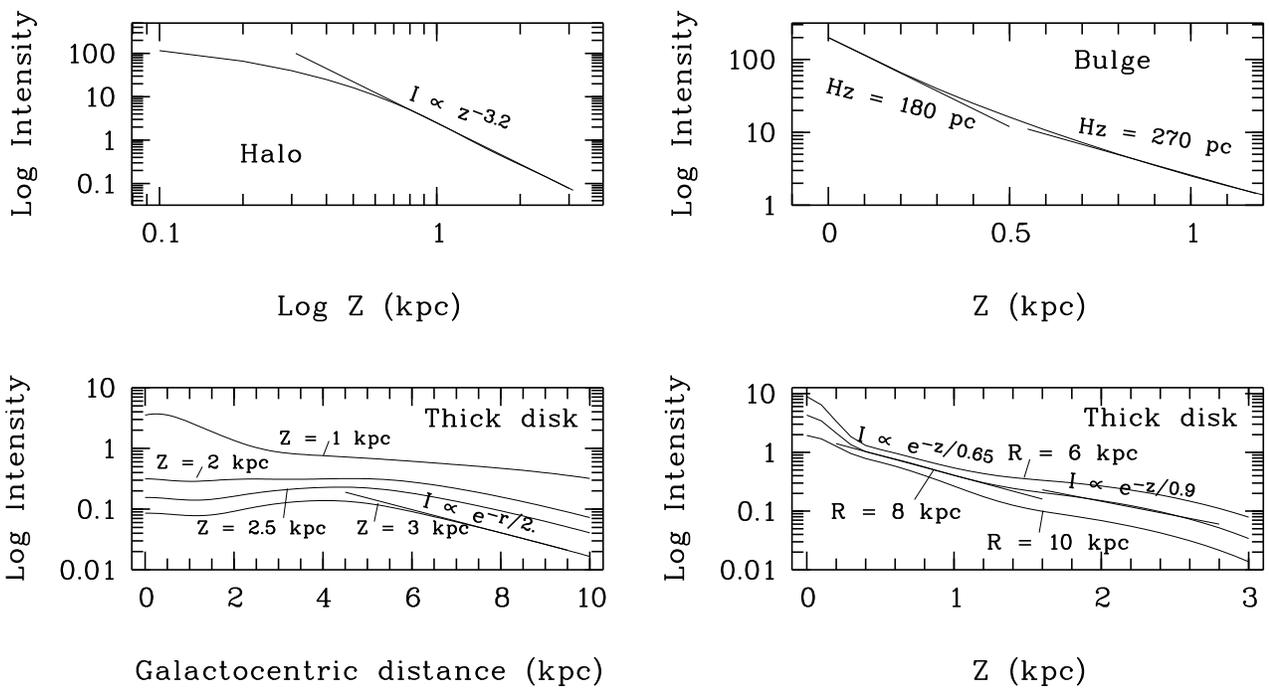

**Fig. 7.** Cuts of the deprojected COBE map in some galactic regions. The derived scale parameters are indicated.

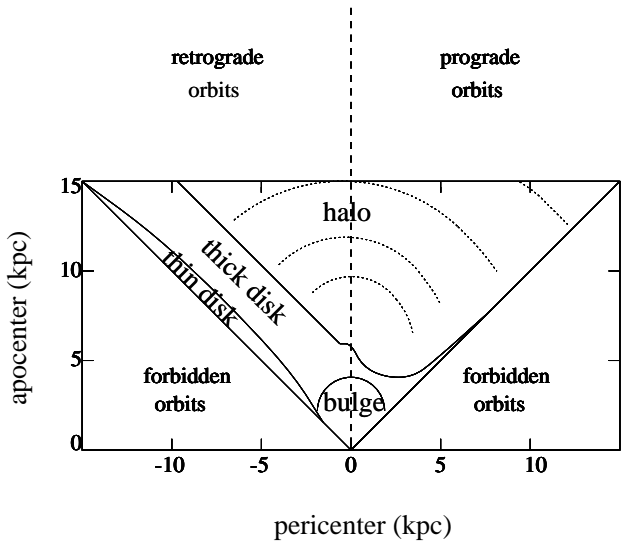

**Fig. 6.** The distribution function in 'turning point' space

### 4.3. The COBE map deprojection

Our dynamical modeling also provides a method for deprojecting an observed light distribution. It suffices to weigh only lightly the kinematic data in the $\chi^2$ variable. The main advantage of such a procedure is that we then know that the deprojected light distribution can be produced by *orbits*, i.e., stems from a positive distribution function. The deprojection, shown in figure 8 (bottom), assumes a galactocentric distance of 8 kpc. The bulge and old disk shapes are very clearly present.

First, let's consider the bulge, the density of which is mainly concentrated inside 0.8 kpc. The radial profile is exponential (except for very small radii, of course), with a scale length $H_r^b = 0.4$ kpc. The determination of the vertical scale height is more problematic because of the uncertainty in the correction of the interstellar extinction : it varies smoothly from $H_z^b = 180$ pc at $Z = 0$ kpc to $H_z^b = 270$ pc at $Z = 0.8$ kpc (see figure 7). If we take 235 pc as the average value of $H_z^b$, then the corresponding $\frac{Z_0}{R_0}$ bulge scale height ratio is 0.56, in very good agreement with other axisymmetric bulge axis ratios values (0.6 in Weiland *et al.* (1994), who use the ratio of the average projected 2.2 $\mu$m COBE intensities, or 0.53 in Dwek *et al.* (1995), who use a Gaussian-type fit on the observed 2.2 $\mu$m COBE map, or again 0.61 in Kent *et al.* (1991), using their oblate spheroid model), but lower in general than the ratios in triaxial bulge models.

The bulge outer region at large $z$-height, which probably forms a smooth transition towards the halo, clearly follows a power law of index 3.2 (figure 7).

As to the disk, the resulting exponential scale length is 2.6 kpc (at the bottom of figure 8). Here there seems to be an obvious chromatic dependence: 2 - 3 kpc in near IR, 3.5 - 5.5 kpc in optical and 4.5 - 6 kpc for IRAS OH/IR stars (Kent *et al.*, 1991). Kinematical determinations show also a large spread but generaly favour a larger scale length (4.4 kpc for Lewis & Freeman, 1989 ; 4.3 - 5.8 kpc for van der Kruit, 1989). On the other hand, our small value is in agreement with recent determinations suggesting that the disk scale length is indeed rather short: Robin *et al.*, (1992), have found $H_r^d = 2.5$ kpc using UBV photometry toward the anticentre down to m = 25 ; Fux & Martinet, (1994), have resolved the asymmetric drift equation by substituting the velocity dispersion gradients with analytical approximations and an application to old disk kinematical data in the solar neighbourhood (with a constant $H_z^d$) has lead them to a value of 2.5 kpc.

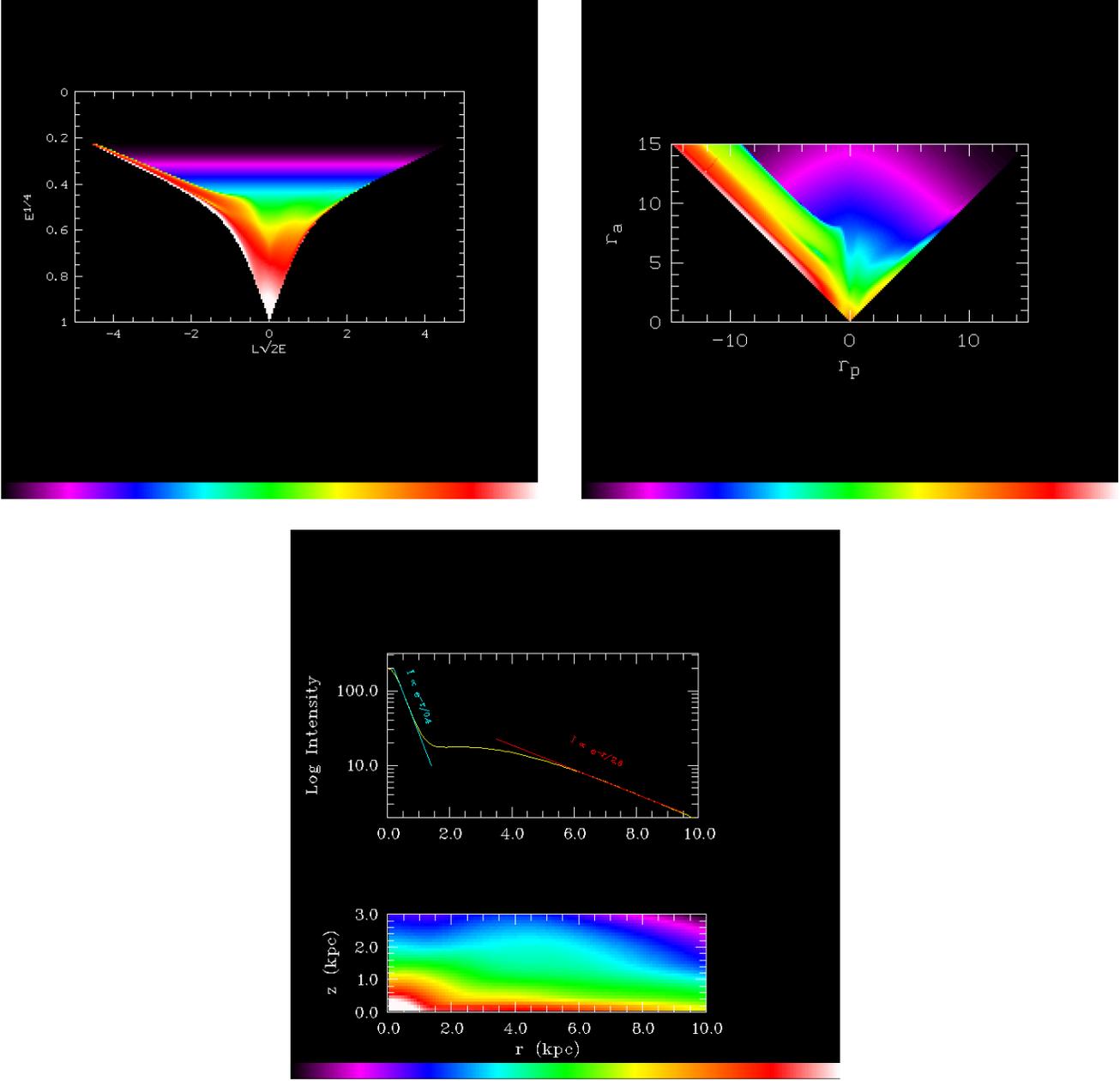

**Fig. 8.** At the top: logarithmic-scaled images of the distribution function in integral space (right) and in turning point space (left). Black indicates lower intensity and white indicates higher intensity.
At the bottom: The 2-integral deprojection of COBE map, presented in logarithmic density scale.

The exponential scale height $H_z^d$ given by the model is constant and equal to 150 pc up to $r = 5.5$ kpc, and increases monotonically beyond that radius, with a gradient of about 20 pc/kpc, to attain for example 210 pc in the solar neighbourhood. This gradient is quite similar to the value found originally by Kent *et al.* (1991), and its presence has been recently suggested again (Fux & Martinet, 1994 ; Ibata & Gilmore, 1995). As to the small value of the scale height, it may point to the fact that the 2.2 $\mu$m emission detected by COBE does not solely originate from an old population, but that a considerable fraction of it actually stems from young disk populations.

There is another (non negligible) disk-like component in our configuration space density map at $z > 0.3$ kpc. It presents all the characteristics of a thick disk, which was originally proposed by Gilmore *et al.*, (1983), and is thought to be populated by intermediate populations. Besides, we have attempted to make a global estimation of the averaged ages of the disks objects by using the relation between ages and dispersion provided by Wielen (1977). The corresponding 2-integral deprojection of the dispersions is shown in figure 9, and show up the possibility that PNe in the thick disk may be estimated to be around 7-9 Gyrs old, contrary to the young disk ones where averaged ages may turn around 2-5 Gyrs.

The radial scale length of the thick disk here appears to be consistent with that of the old disk, namely 2 kpc. Vertical scaleheights are many, with typical values varying between 0.5 and 1 kpc (figure 7).

Finally, at about $z = 4$ kpc a transition between the thick disk and halo occurs, which is probably not well determined by our 2-integral model.

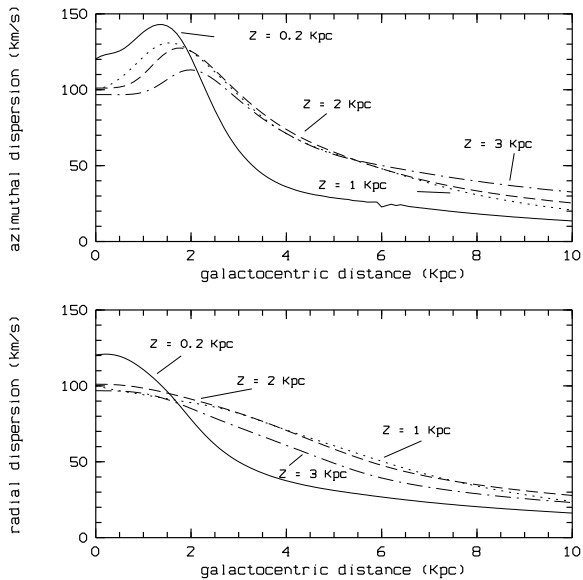

**Fig. 9.** Radial (up figure) and tangential (down figure) dispersions of PNe at Z = 0.2, 1, 2 and 3 kpc, as provided by the 2-integral deprojection.

## 5. Concluding remarks

The extinction-corrected 2.2 $\mu$m COBE map proves to be a very useful substitute for AGB and post-AGB star counts in the presence of non-kinematic detection biases. It can also be used to isolate in the models the effects of kinematics alone, and one could thus compare the dynamics of different samples if one uses systematically the same projected density map for all of them.

The bulge-disk decomposition is very obvious in the density map in $(r, z)$ obtained via the 2-integral deprojection of the COBE data, which allows us to estimate the scale parameters of all the principal galactic components: for example, our model conforts some recent ideas about the disk as a scale-length rather short (2.6 kpc) and a vertical profile different from 'pure-disk' galaxies ones where scale heights are independent of galactocentric distance. There is also evidence that the 2.2 $\mu$m disk emission has a considerable component that originates from young populations.

This is the very first time that a dynamical model with a global distribution function is made of a PNe sample.

An important conclusion of our paper remains that a 2-integral model seems not really adequate to characterize the dynamical state of PNe, thereby demonstrating the possible presence and importance of the third integral on a global scale. Of course, the dependence of the distribution function on the third integral in the solar neigbourhood has been established a long time ago, and it is widely believed on firm theoretical grounds that this is also the case on a global scale. The observational demonstration is not trivial however, since the effect of the third integral is mostly felt in the vertical velocity, which contributes only weakly to the radial velocities of objects in the disk due the position of the sun. Global models of the kind presented in this paper are able to exclude 2-integral models, because for such models $\sigma_r = \sigma_z$, $\sigma_r$ is well observable in the radial velocities and both dispersions are uniquely determined by the morphology of the stellar component (i.e. the COBE map).

The failure of our 2-integral models is also confirmed by the results of Sevenster et al. (1995), where the same modeling scheme is applied on a sample of 700 galactic OH/IR stars (the sample 'tLH'). If we compare the lower right panel of their figure 4 with the lower right panel of our figure 3, we see a very similar behaviour, though probably a bit less pronounced for the OH/IRs: the velocity dispersion of the models are too low at large longitudes. Sevenster et al. also note in passing that their tLH sample may need an additional integral. This is certainly to be expected, since the kinematics of the OH/IRs and the PNe are very similar, except that the dispersions in the OH/IR population are a bit lower and that the modeling of the OH/IRs used star counts instead of the COBE map because the OH/IR sample is much more homogeneous.

**Acknowledgement** The authors thank the referee, T. de Zeeuw, for his comments which helped improve the paper, and also F. Chambat for his early work on the subject.

The COBE datasets were developed by the NASA Goddard Space Flight center under the guidance of the COBE Science Working Group and were provided by the NSSDC.